\documentclass[10pt,a4paper,english,twocolumn, conference]{IEEEtran}
\usepackage[T1]{fontenc}
\usepackage{verbatim}
\usepackage{amsmath}
\usepackage{graphicx}

\makeatletter

\pdfpageheight\paperheight
\pdfpagewidth\paperwidth

\usepackage{epstopdf}
\usepackage{balance}
\usepackage{cite}
\usepackage{citesort}
\IEEEoverridecommandlockouts

\@ifundefined{showcaptionsetup}{}{%
 \PassOptionsToPackage{caption=false}{subfig}}
\usepackage{subfig}
\makeatother

\usepackage{babel}
\begin{document}

\title{Promises and Caveats of Uplink\\
IoT Ultra-Dense Networks
}

\author{\noindent {\normalsize{}Ming Ding$^{\ddagger}$, David L$\acute{\textrm{o}}$pez
P$\acute{\textrm{e}}$rez$^{\dagger}$}\emph{\normalsize{}}\\
\textit{\small{}$^{\ddagger}$Data61, Australia }{\small{}\{Ming.Ding@data61.csiro.au\}}\textit{\small{}}\\
\textit{\small{}$^{\dagger}$Nokia Bell Labs, Ireland }{\small{}\{david.lopez-perez@nokia.com\}}
}
\maketitle
\begin{abstract}
In this paper, by means of simulations, we evaluate the uplink (UL)
performance of an Internet of Things (IoT) capable ultra-dense network
(UDN) in terms of the coverage probability and the density of reliably
working user equipments (UEs). From our study, we show the benefits
and challenges that UL IoT UDNs will bring about in the future. In
more detail, for a low-reliability criterion, such as achieving a
UL signal-to-interference-plus-noise ratio (SINR) above 0$\,$dB,
the density of reliably working UEs grows quickly with the network
densification, showing the potential of UL IoT UDNs. In contrast,
for a high-reliability criterion, such as achieving a UL SINR above
10$\,$dB, the density of reliably working UEs remains to be low in
UDNs due to excessive inter-cell interference, which should be considered
when operating UL IoT UDNs. Moreover, considering the existence of
a non-zero antenna height difference between base stations (BSs) and
UEs, the density of reliably working UEs could even \emph{decrease}
as we deploy more BSs. This calls for the usage of sophisticated interference
management schemes and/or beam steering/shaping technologies in UL
IoT UDNs. 
\end{abstract}

\section{Introduction\label{sec:Introduction}}

Recent years have seen rapid advancement in the development and deployment
of Internet of Things (IoT) networks, which can be attributed to the
increasing communication and sensing capabilities combined with the
falling prices of IoT devices~\cite{Al-Fuqaha2015IoTsurvey}. For
example, base stations (BSs) can be equipped with the latest IoT technologies,
such as the new generation of machine type communications~\cite{Tutor_smallcell},
to collect data from gas, water, and power meters via uplink (UL)
transmissions. In practice, such BSs can take the form of both terrestrial
and aerial ones~\cite{Fotouhi2017UAVIoT}. 

This poses, however, a challenge to the wireless industry, which must
offer an increasing volume of reliable traffic in a profitable and
energy efficient manner, especially for the UL communications.%
{} In this context, the orthogonal deployment of ultra-dense (UD) small
cell networks (SCNs), or simply ultra-dense networks (UDNs), have
been selected as one of the workhorse for network enhancement in the
fourth-generation (4G) and fifth-generation (5G) networks developed
by the 3rd Generation Partnership Project (3GPP)~\cite{Tutor_smallcell}.
Here, orthogonal deployment means that UDNs and macrocell networks
operate on different frequency spectrum, which simplifies network
management due to no inter-tier interference. 

The performance analysis of IoT UDNs is, however, particularly challenging
for the UL because \emph{i) }UDNs are fundamentally different from
the current sparse/dense networks~\cite{Ding2017UDNmag} and \emph{ii)
}the UL power control mechanism operates according to the random user
equipment (UE) positions in the network, which is quite different
from the constant power setting in the downlink (DL)~\cite{Our_DNA_work_TWC15}.

In this paper, by means of simulations, we evaluate the network performance
of UL IoT UDNs in terms of the coverage probability and the density
of reliably working UEs. The main findings of this paper are as follows: 
\begin{itemize}
\item We find that for a low-reliability criterion, such as achieving a
UL signal-to-interference-plus-noise ratio (SINR) above $\gamma=0\,\mathrm{dB}$,
the density of reliably working UEs quickly grows with the network
densification, showing the benefits of UL IoT UDNs. In contrast, for
a high-reliability criterion, such as achieving a UL SINR above $\gamma=10\,\mathrm{dB}$,
the density of reliably working UEs remains low in UDNs due to excessive
inter-cell interference, which should be considered when operating
UL IoT UDNs. 
\item We find that due to the existence of a non-zero antenna height difference
between BSs and UEs, the density of reliably working UEs could even
\emph{decrease} as we deploy more BSs in a UL IoT UDN. This calls
for the usage of sophisticated interference management schemes and/or
beam steering/shaping technologies in UL IoT UDNs. 
\item We find that the correlated shadow fading allows a BS with a lower
environmental fading factor to provide connection to a larger number
of UEs. Thus, its theoretical analysis is an open problem for further
study. 
\item We find that the optimized hexagonal-like BS deployment can improve
network performance for relatively sparse networks, but not for UDNs.
Thus, its theoretical study is not urgent. 
\end{itemize}

\section{Discussion on the Assumptions of UL IoT UDNs \label{sec:Discussion-NA}}

In this section, we discuss several important assumptions in UL IoT
UDNs. 

\subsection{BS Deployments\label{subsec:BS-Deployments}}

\begin{figure*}
\center\subfloat[\label{fig:fullCov_hexa}The hexagonal BS deployment.]{\includegraphics[width=5cm]{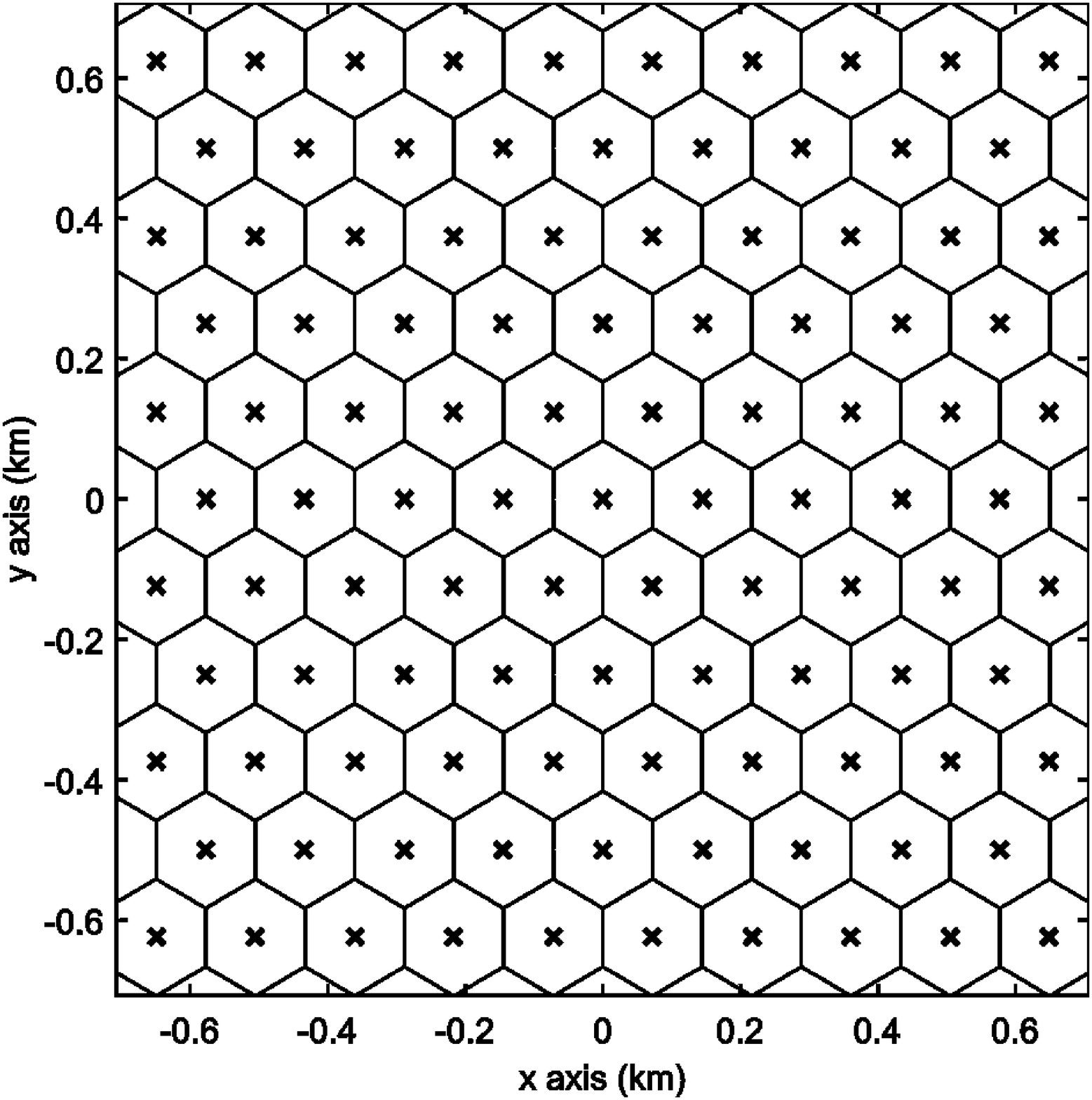}\vspace{-0.1cm}
}$\quad\quad\quad\quad\quad\quad\quad\quad$\subfloat[\label{fig:fullCov_rand}The random BS deployment.]{\includegraphics[width=5cm]{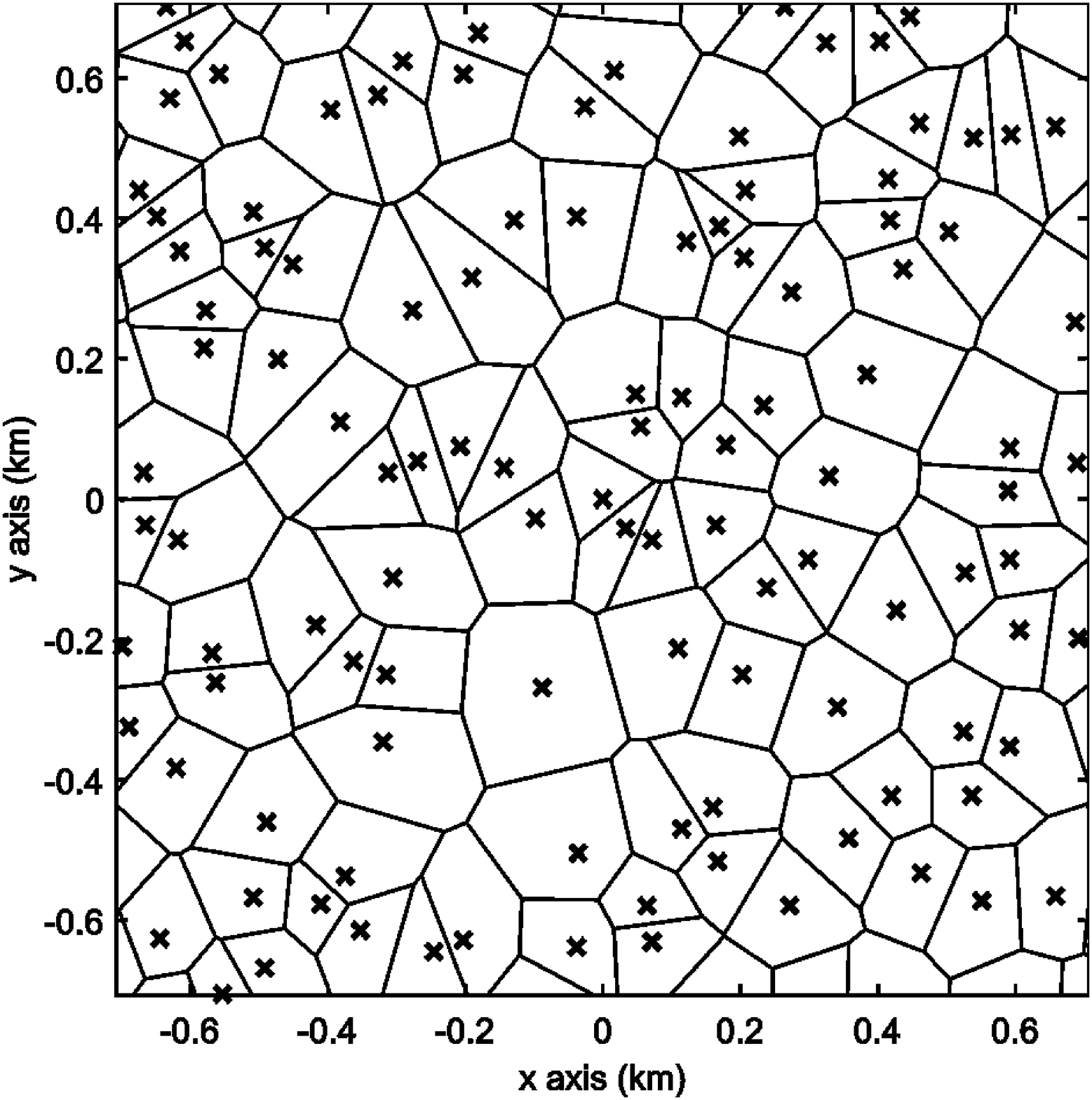}\vspace{-0.1cm}
}\renewcommand{\figurename}{Fig.}\caption{\label{fig:illus_4_NSs}Illustration of two widely accepted types
of BS deployments. Here, BSs are represented by markers ``x'' and
cell coverage areas for user equipment (UE) distribution are outlined
by solid lines.}
\vspace{-0.3cm}
\end{figure*}

In general, to study the performance of a UL IoT network, two types
of BS deployments can be found in the literature, i.e., the hexagonal
and the random deployments as shown in Fig.~\ref{fig:illus_4_NSs}.

In Fig.~\ref{fig:illus_4_NSs}, BSs are represented by markers ``x''
and cell coverage areas are outlined by solid lines. Note that the
hexagonal BS deployment leads to an upper-bound performance because
BSs are evenly distributed in the network scenario, and thus very
strong interference due to close BS proximity is precluded~\cite{Jeff2011}.
In contrast, the random BS deployment reflects a more realistic network
deployment with more challenging interference conditions~\cite{Tutor_smallcell,TR36.872}.
For completeness, in our following performance evaluation, we will
consider both BS deployments. 

\subsection{Antenna Heights\label{subsec:Antenna-Heights}}

In the performance analysis of the conventional sparse or dense cellular
networks, the antenna height difference between BSs and UEs is usually
ignored due to the dominance of the horizontal distance. However,
with a much shorter distance between a BS and its served UEs in an
UDN, such antenna height difference becomes non-negligible~\cite{TR36.872}.
The performance impact of such antenna height difference between BSs
and UEs on the DL UDNs has been investigated in~\cite{Ding2016ASECrash}.
More specifically, the existence of a non-zero antenna height difference
between BSs and UEs gives rise to a non-zero cap on the minimum distance
between a BS and its served UEs, and thus \emph{a cap on the received
signal power strength}. Thus, and although each inter-cell interference
power strength is subject to the same cap, the aggregate inter-cell
interference power will overwhelm the signal power in an UDN due to
the sheer number of strong interferers. Consequently, the antenna
height difference between BSs and UEs should be considered in the
performance evaluation of UL IoT UDNs. 

\subsection{Line-of-Sight Transmissions\label{subsec:Line-of-Sight-Transmissions}}

A much shorter distance between a BS and its served UEs in an UDN
implies a higher probability of line-of-sight (LoS) transmissions.
The performance impact of such LoS transmissions on the DL of UDNs
has been shown to be significant in~\cite{related_work_Jeff,our_work_TWC2016}.
Generally speaking, LoS transmissions are more helpful to enhance
the received signal strength than non-line-of-sight (NLoS) transmissions.
However, after a certain level of BS densification, not only the signal
power, but also the inter-cell interference power will significantly
grow due to \emph{the emergence of LoS interfering paths}. Thus, the
transition of a large number of interfering paths from NLoS to LoS
will overwhelm the signal power in an UDN. Consequently, the probabilistic
LoS transmissions should also be considered in the performance evaluation
of UL IoT UDNs. 

\subsection{BS Idle Mode\label{subsec:BS-Idle-Mode}}

Considering the surplus of BSs in UDNs, a BS can be put to sleep when
there is no active UE connected to it, which is referred as the BS
idle mode capability (IMC)~\cite{dynOnOff_Huang2012,Ding2016TWC_IMC}.
As a result, the active UEs' SINR can benefit from \emph{i)} a BS
selection diversity gain, i.e., each UE has a plurality of BSs to
select its server from, and \emph{ii) }a tight control of inter-cell
interference, as effective UL inter-cell interference only comes from
active UEs served by active neighboring BSs. The surplus of BSs together
with the IMC can be seen as a powerful tool that can mitigate the
interference problems presented in the previous subsections. However,
it should be noted that switching off BSs has a negative impact on
the number of IoT UEs that can concurrently transmit.

\section{System Model\label{sec:System-Model}}

For a certain time-frequency resource block (RB), we consider a UL
IoT network with BSs deployed on a plane according to the hexagonal
deployment or the random deployment, as shown in Fig.~\ref{fig:illus_4_NSs}.
For both BS deployments, the density of BSs is denoted by $\lambda$
$\textrm{BSs/km}^{2}$. Furthermore, we consider a homogeneous Poisson
point process (HPPP) $\Phi$ to characterize the random deployment. 

Active UEs are assumed to be distributed following an HPPP with a
density of $\rho$ $\textrm{UEs/km}^{2}$. Here, we only consider
active UEs in the network because non-active UEs do not trigger data
transmission. Note that the total number of UEs, e.g., phones, gateways,
sensors, tags, etc., in a UL IoT network should be much higher than
the number of the active UEs. However, we believe that in a certain
time-frequency RB, the active UEs with non-zero data traffic demands
should still be not many. For example, a typical density of active
UEs is around $300\thinspace\textrm{UEs/km}^{2}$ in 5G~\cite{Tutor_smallcell}.

\subsection{Channel Model\label{subsec:Channel-Model}}

The two-dimensional (2D) distance between a BS and a UE is denoted
by $r$. Moreover, the absolute antenna height difference between
a BS and a UE is denoted by $L$. Thus, the 3D distance between a
BS and a UE can be expressed as
\begin{equation}
w=\sqrt{r^{2}+L^{2}}.\label{eq:actual_dis_BS2UE}
\end{equation}
Note that the value of $L$ is in the order of several meters~\cite{TR36.828}.

Following~\cite{our_work_TWC2016}, we adopt a general path loss
model, where the path loss $\zeta\left(w\right)$ is a multi-piece
function of $w$ written as%
\begin{equation}
\zeta\left(w\right)=\begin{cases}
\zeta_{1}\left(w\right), & \textrm{when }L\leq w\leq d_{1}\\
\zeta_{2}\left(w\right), & \textrm{when }d_{1}<w\leq d_{2}\\
\vdots & \vdots\\
\zeta_{N}\left(w\right), & \textrm{when }w>d_{N-1}
\end{cases},\label{eq:prop_PL_model}
\end{equation}
where each piece $\zeta_{n}\left(w\right),n\in\left\{ 1,2,\ldots,N\right\} $
is modeled as
\begin{equation}
\zeta_{n}\left(w\right)\hspace{-0.1cm}=\hspace{-0.1cm}\begin{cases}
\hspace{-0.2cm}\begin{array}{l}
\zeta_{n}^{{\rm {L}}}\left(w\right)=A_{n}^{{\rm {L}}}w^{-\alpha_{n}^{{\rm {L}}}},\\
\zeta_{n}^{{\rm {NL}}}\left(w\right)=A_{n}^{{\rm {NL}}}w^{-\alpha_{n}^{{\rm {NL}}}},
\end{array} & \hspace{-0.2cm}\hspace{-0.3cm}\begin{array}{l}
\textrm{LoS:}~\textrm{Pr}_{n}^{{\rm {L}}}\left(w\right)\\
\textrm{NLoS:}~1-\textrm{Pr}_{n}^{{\rm {L}}}\left(w\right)
\end{array}\hspace{-0.1cm}\hspace{-0.1cm},\hspace{-0.1cm}\hspace{-0.1cm}\end{cases}\label{eq:PL_BS2UE}
\end{equation}
where
\begin{itemize}
\item $\zeta_{n}^{{\rm {L}}}\left(w\right)$ and $\zeta_{n}^{{\rm {NL}}}\left(w\right),n\in\left\{ 1,2,\ldots,N\right\} $
are the $n$-th piece path loss functions for the LoS and the NLoS
cases, respectively,
\item $A_{n}^{{\rm {L}}}$ and $A_{n}^{{\rm {NL}}}$ are the path losses
at a reference 3D distance $w=1$ for the LoS and the NLoS cases,
respectively,
\item $\alpha_{n}^{{\rm {L}}}$ and $\alpha_{n}^{{\rm {NL}}}$ are the path
loss exponents for the LoS and the NLoS cases, respectively, and
\item $\textrm{Pr}_{n}^{{\rm {L}}}\left(w\right)$ is the $n$-th piece
LoS probability function that a transmitter and a receiver separated
by a 3D distance $w$ has an LoS path, which is assumed to be \emph{a
monotonically decreasing function} with respect to $w$. Existing
measurement studies have confirmed this assumption~\cite{TR36.828}.
\end{itemize}
Moreover, we assume that each BS/UE is equipped with an isotropic
antenna, and that the multi-path fading between a BS and a UE is independently
identical distributed (i.i.d.) Rayleigh distributed~\cite{related_work_Jeff,Related_work_Health,our_work_TWC2016}. 

\subsection{User Association Strategy\label{subsec:User-Association-Strategy}}

We assume a practical user association strategy (UAS), in which each
UE is connected to the BS giving the maximum average received signal
strength~\cite{Related_work_Health,our_work_TWC2016}.%
{} Such UAS can be formulated by
\begin{equation}
b_{o}=\underset{b}{\arg\max}\left\{ \bar{R}_{b}\left(w\right)\right\} ,\label{eq:UAS_smallestPL}
\end{equation}
where $\bar{R}_{b}\left(w\right)$ denotes the average received signal
strength from BS $b$ and the UE of interest, separated by a distance
of $w$. Assuming a constant BS transmission power, $\bar{R}_{b}\left(w\right)$
can be equivalently evaluated by $\zeta\left(w\right)$ defined in
(\ref{eq:prop_PL_model}).

As a special case to show our numerical results in the simulation
section, we consider a practical two-piece path loss function and
a two-piece exponential LoS probability function, defined by the 3GPP~\cite{TR36.828}.
More specifically, in (\ref{eq:prop_PL_model}) we use $N=2$, $\zeta_{1}^{{\rm {L}}}\left(w\right)=\zeta_{2}^{{\rm {L}}}\left(w\right)=A^{{\rm {L}}}w^{-\alpha^{{\rm {L}}}}$,
$\zeta_{1}^{{\rm {NL}}}\left(w\right)=\zeta_{2}^{{\rm {NL}}}\left(w\right)=A^{{\rm {NL}}}w^{-\alpha^{{\rm {NL}}}}$,
$\textrm{Pr}_{1}^{{\rm {L}}}\left(w\right)=1-5\exp\left(-R_{1}/w\right)$,
and $\textrm{Pr}_{2}^{{\rm {L}}}\left(w\right)=5\exp\left(-w/R_{2}\right)$,
where $R_{1}=156$\ m, $R_{2}=30$\ m, and $d_{1}=\frac{R_{1}}{\ln10}=67.75$\ m~\cite{TR36.828}.
For clarity, this path loss case is referred to as \textbf{the 3GPP
Case} hereafter.

\subsection{BS Activation Model\label{subsec:BS-Activation-Model}}

As discussed in Subsection~\ref{subsec:BS-Idle-Mode}, a BS will
enter an idle mode if there is no UE connected to it. Thus, the set
of active BSs is determined by the UAS. Since UEs are randomly and
uniformly distributed in the network and given the adopted UAS strategy,
we can assume that the active BSs also follow an HPPP distribution
$\tilde{\Phi}$~\cite{dynOnOff_Huang2012}, the density of which
is denoted by $\tilde{\lambda}$ $\textrm{BSs/km}^{2}$, where $\tilde{\lambda}\leq\lambda$
and $\tilde{\lambda}\leq\rho$. Note that $\tilde{\lambda}$ also
characterizes the density of active UEs because no collision exists
in the centralized cellular IoT UDNs.%

For illustration purposes, considering a single-slope path loss model
and a nearest-BS UAS, $\tilde{\lambda}$ can be calculated as~\cite{dynOnOff_Huang2012}%
\begin{equation}
\tilde{\lambda}=\lambda\left[1-\frac{1}{\left(1+\frac{\rho}{q\lambda}\right)^{q}}\right],\label{eq:lambda_tilde_Huang}
\end{equation}
where an empirical value of 3.5 was suggested for $q$ in~\cite{dynOnOff_Huang2012}\footnote{Note that according to~\cite{Ding2016TWC_IMC}, $q$ should also
depend on the path loss model with LoS/NLoS transmissions. Having
said that, \cite{Ding2016TWC_IMC} also showed that (\ref{eq:lambda_tilde_Huang})
is generally very accurate to characterize $\tilde{\lambda}$ for
the 3GPP Case with LoS/NLoS transmissions.}.

\subsection{UL Power Control Model}

The UE power, denoted by $P^{\textrm{UE}}$, is subject to semi-static
power control (PC) in practice. In this paper, we adopt the fractional
path loss compensation (FPC) scheme standardized in 4G~\cite{TR36.828},
which can be modeled as 
\begin{equation}
P^{{\rm {UE}}}=10^{\frac{P_{0}}{{10}}}\left[\zeta\left(w\right)\right]^{-\eta}N^{{\rm {RB}}},\label{eq:UL_P_UE}
\end{equation}
where $P_{0}$ is the target received power in dBm on each RB at the
BS, $\eta\in\left(0,1\right]$ is the FPC compensation factor, and
$N^{{\rm {RB}}}$ is the number of RBs in the frequency domain. 

\subsection{The Coverage Probability\label{subsec:The-Coverage-Probability}}

Based on this system model, we can define the coverage probability
that the typical UE's UL SINR is above a designated threshold $\gamma$
as
\begin{equation}
p^{{\rm {cov}}}\left(\lambda,\gamma\right)=\textrm{Pr}\left[\mathrm{SINR^{{\rm {U}}}}>\gamma\right],\label{eq:Coverage_Prob_def}
\end{equation}
where the UL SINR is calculated by
\begin{equation}
\mathrm{SINR}^{{\rm {U}}}=\frac{P_{b_{o}}^{{\rm {UE}}}\zeta\left(w_{b_{o}}\right)h}{I_{{\rm {agg}}}^{{\rm {U}}}+P_{{\rm {N}}}^{{\rm {U}}}},\label{eq:SINR}
\end{equation}
where $b_{o}$ denotes the serving BS of the typical UE, $P_{b_{o}}^{{\rm {UE}}}$
is the UE transmission power given by (\ref{eq:UL_P_UE}), $w_{b_{o}}$
is the distance from the typical UE to its serving BS $b_{o}$, $h$
is the Rayleigh channel gain modeled as an exponentially distributed
random variable (RV) with a mean of one as mentioned above, $P_{{\rm {N}}}^{{\rm {U}}}$
is the additive white Gaussian noise (AWGN) power at the serving BS
$b_{o}$, and $I_{{\rm {agg}}}^{{\rm {U}}}$ is the UL aggregate interference.%

\subsection{The Density of Reliably Working UEs}

Based on the definitions of the active BS density in Subsection~\ref{subsec:BS-Activation-Model}
and the coverage probability in Subsection~\ref{subsec:The-Coverage-Probability},
we can further define a density of reliably working UEs that can operate
above a target UL SINR threshold $\gamma$ as

\begin{equation}
\tilde{\rho}=\tilde{\lambda}p^{{\rm {cov}}}\left(\lambda,\gamma\right),\label{eq:reliable_UE_den_def}
\end{equation}
where the active BS density $\tilde{\lambda}$ measures the maximum
density of UEs that can simultaneously transmit, and the coverage
probability $p^{{\rm {cov}}}\left(\lambda,\gamma\right)$ scales down
$\tilde{\lambda}$, giving the density of reliably working UEs. The
larger the UL SINR threshold $\gamma$, the higher the reliability
of the IoT communications, and thus the less the UEs that can simultaneously
achieve such reliability.

\subsection{More Refined Assumptions}

In performance analysis, the multi-path fading is usually modeled
as Rayleigh fading for simplicity. However, in the 3GPP, a more practical
model based on generalized Rician fading is widely adopted for LoS
transmissions~\cite{SCM_pathloss_model}. In such model, the $K$
factor in dB scale (the ratio between the power in the direct path
and the power in the other scattered paths) is modeled as $K[\textrm{dB}]=13-0.03w$,
where $w$ is defined in (\ref{eq:actual_dis_BS2UE}). More specifically,
let $h^{{\rm {L}}}$ denote the multi-path fading power for LoS transmissions.
Then, for the 3GPP model of Rician fading, $h^{{\rm {L}}}$ follows
a non-central chi-squared distribution with its PDF given by~\cite{Book_Integrals}

\noindent 
\begin{eqnarray}
f\left(h^{{\rm {L}}}\right) & = & \left(K+1\right)\exp\left(-K-\left(K+1\right)h^{{\rm {L}}}\right)\nonumber \\
 &  & \times I_{0}\left(2\sqrt{K\left(K+1\right)h^{{\rm {L}}}}\right),\label{eq:PDF_Rician}
\end{eqnarray}
where $K$ is the distance-dependent value discussed above and $I_{0}\left(\cdot\right)$
is the 0-th order modified Bessel function of the first kind~\cite{Book_Integrals}. 

Moreover, the shadow fading is also usually not considered or simply
modeled as i.i.d. RVs in performance analysis. However, in the 3GPP,
a more practical correlated shadow fading is often used~\cite{TR36.828,SCM_pathloss_model,Ding2015corrSF},
where the shadow fading in dB unit is modeled as zero-mean Gaussian
RV~\cite{TR36.828}. More specifically, the shadow fading coefficient
in dB unit between BS $b$ and UE $u$ is formulated as~\cite{TR36.828}
\begin{equation}
S_{bu}=\sqrt{\tau}S_{u}^{{\rm {UE}}}+\sqrt{1-\tau}S_{b}^{{\rm {BS}}},\label{eq:corr_shad_dB}
\end{equation}

\noindent where $\tau$ is the correlation coefficient of shadow fading,
$S_{u}^{{\rm {UE}}}$ and $S_{b}^{{\rm {BS}}}$ are i.i.d. zero-mean
Gaussian RVs attributable to UE $u$ and BS $b$, respectively. The
variance of $S_{u}^{{\rm {UE}}}$ and $S_{b}^{{\rm {BS}}}$ is denoted
by $\sigma_{\textrm{Shad}}^{2}$. In~\cite{TR36.828}, it is suggested
that $\tau=0.5$ and $\sigma_{\textrm{Shad}}=10\,\mathrm{dB}$.%

Considering the distance-dependent Rician fading for LoS transmissions
and the correlated shadow fading, we can upgrade the 3GPP Case to
an advanced 3GPP Case. In the next section, we will present simulation
results of UL IoT UDNs for both the 3GPP Case and \textbf{the Advanced
3GPP Case}. It should be noted that for the Advanced 3GPP Case, shadow
fading should be considered in the computation of $\bar{R}_{b}\left(w\right)$,
i.e., $\bar{R}_{b}\left(w\right)$ should be evaluated by $\zeta\left(w\right)\times10^{\frac{S_{bu}}{10}}$
in (\ref{eq:UAS_smallestPL}).

\section{Simulation and Discussion\label{sec:Simulation-and-Discussion}}

In this section, we present numerical results to validate the accuracy
of our analysis. According to Tables~A.1-3\textasciitilde{}A.1-7
of~\cite{TR36.828}%
, we adopt the following parameters for the 3GPP Case: $\alpha^{{\rm {L}}}=2.09$,
$\alpha^{{\rm {NL}}}=3.75$, $A^{{\rm {L}}}=10^{-10.38}$, $A^{{\rm {NL}}}=10^{-14.54}$%
, $P_{0}=-76$\ dBm, $\eta=0.8$, $N^{{\rm {RB}}}=55$, $P_{{\rm {N}}}=-91$\ dBm
(with a noise figure of 13\ dB), $\tau=0.5$ and $\sigma_{\textrm{Shad}}=10\,\mathrm{dB}$.%
{} 

\subsection{Performance Results of the 3GPP Case\label{subsec:Performance-Results-3GPP-Case}}

\begin{figure*}
\noindent \begin{centering}
\center\subfloat[\label{fig:lambda_tilde_simp_0dB}The active BS density $\tilde{\lambda}$.]{\includegraphics[width=5cm]{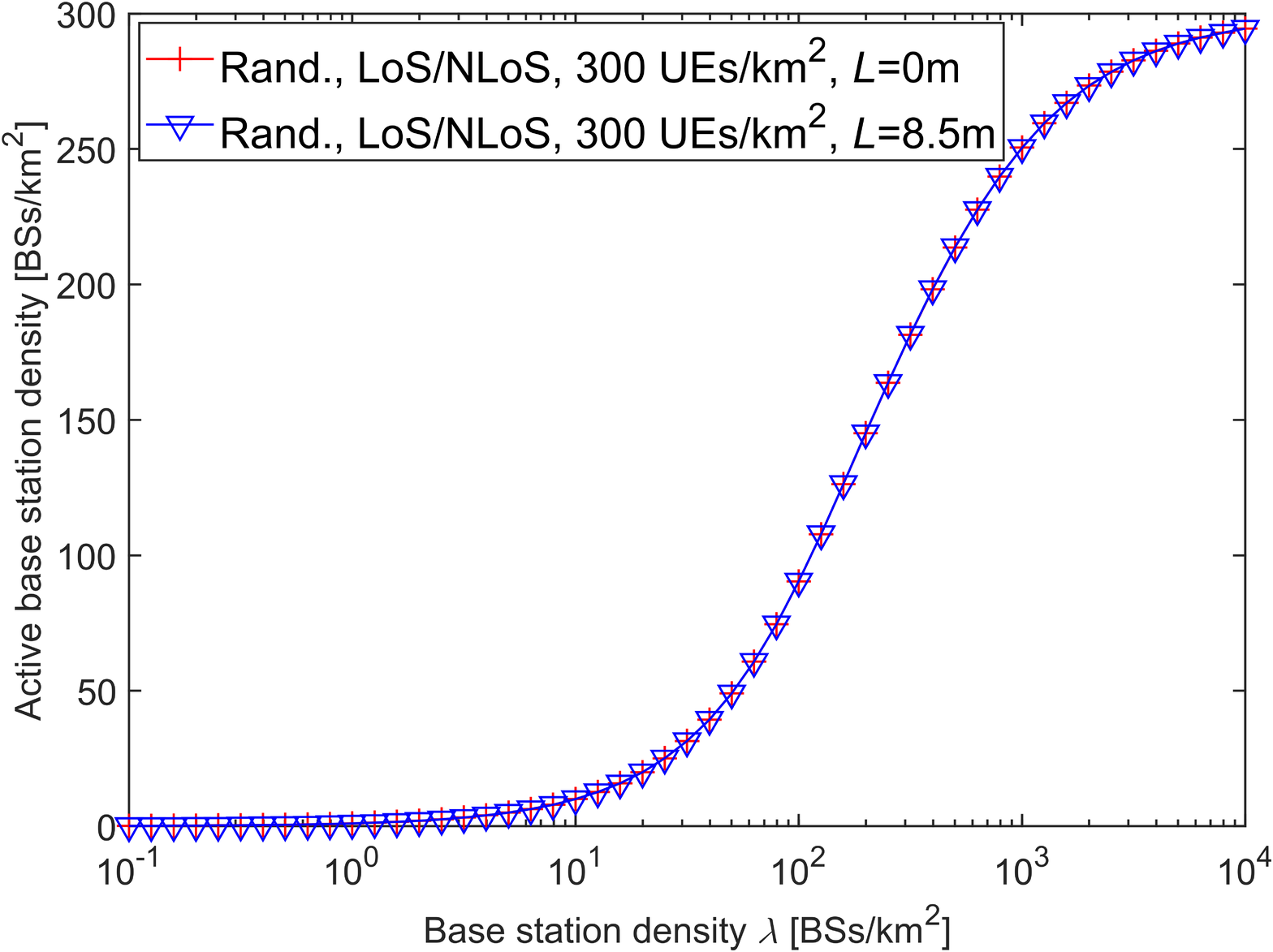}\vspace{-0.1cm}
}$\quad$\subfloat[\label{fig:Pr_cov_simp_0dB}The coverage probability $p^{{\rm {cov}}}\left(\lambda,\gamma\right)$.]{\includegraphics[width=5cm]{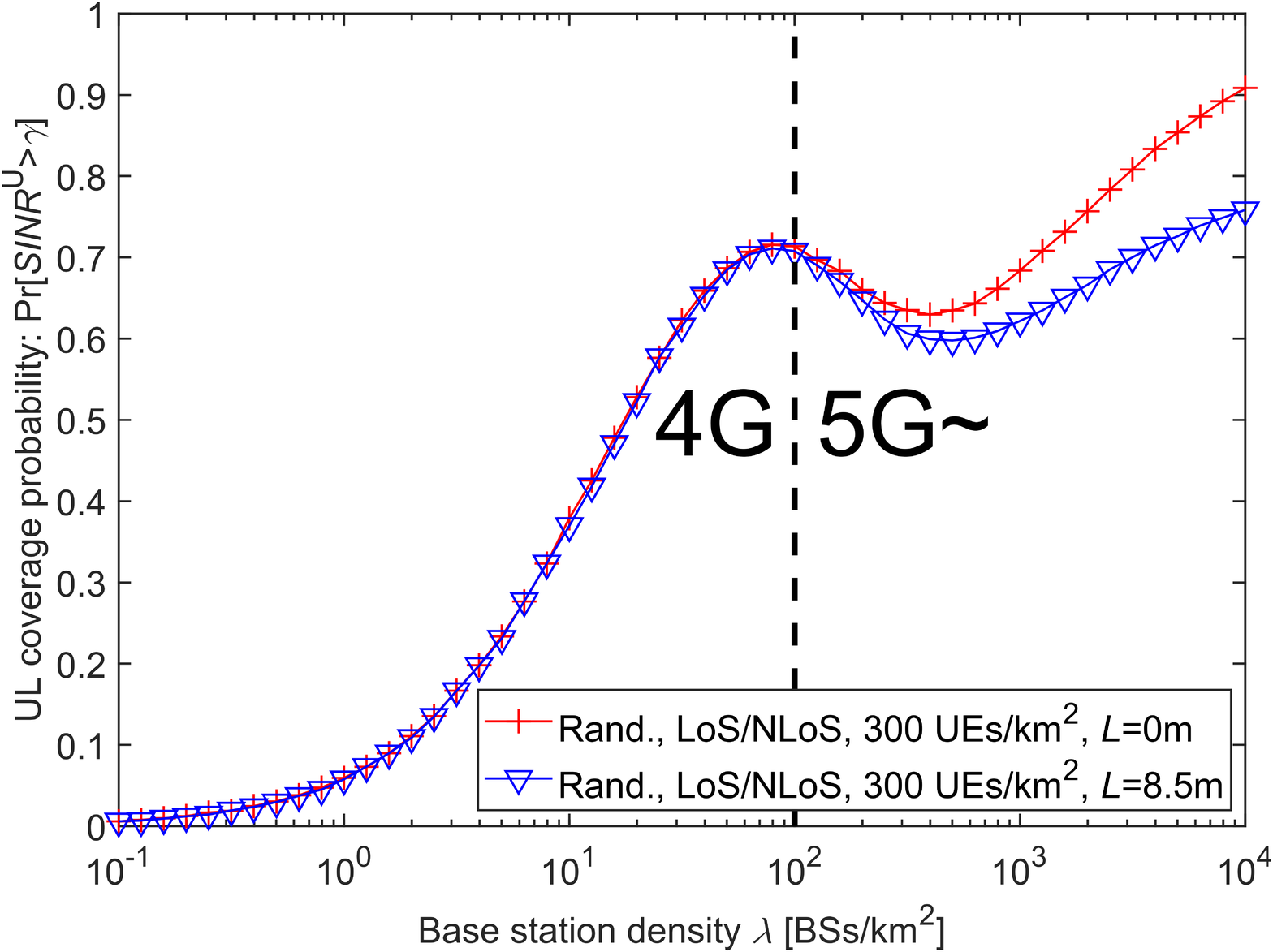}\vspace{-0.1cm}
}$\quad$\subfloat[\label{fig:reliable_UEden_simp_0dB}The density of reliably working
UEs $\tilde{\rho}$.]{\includegraphics[width=5cm]{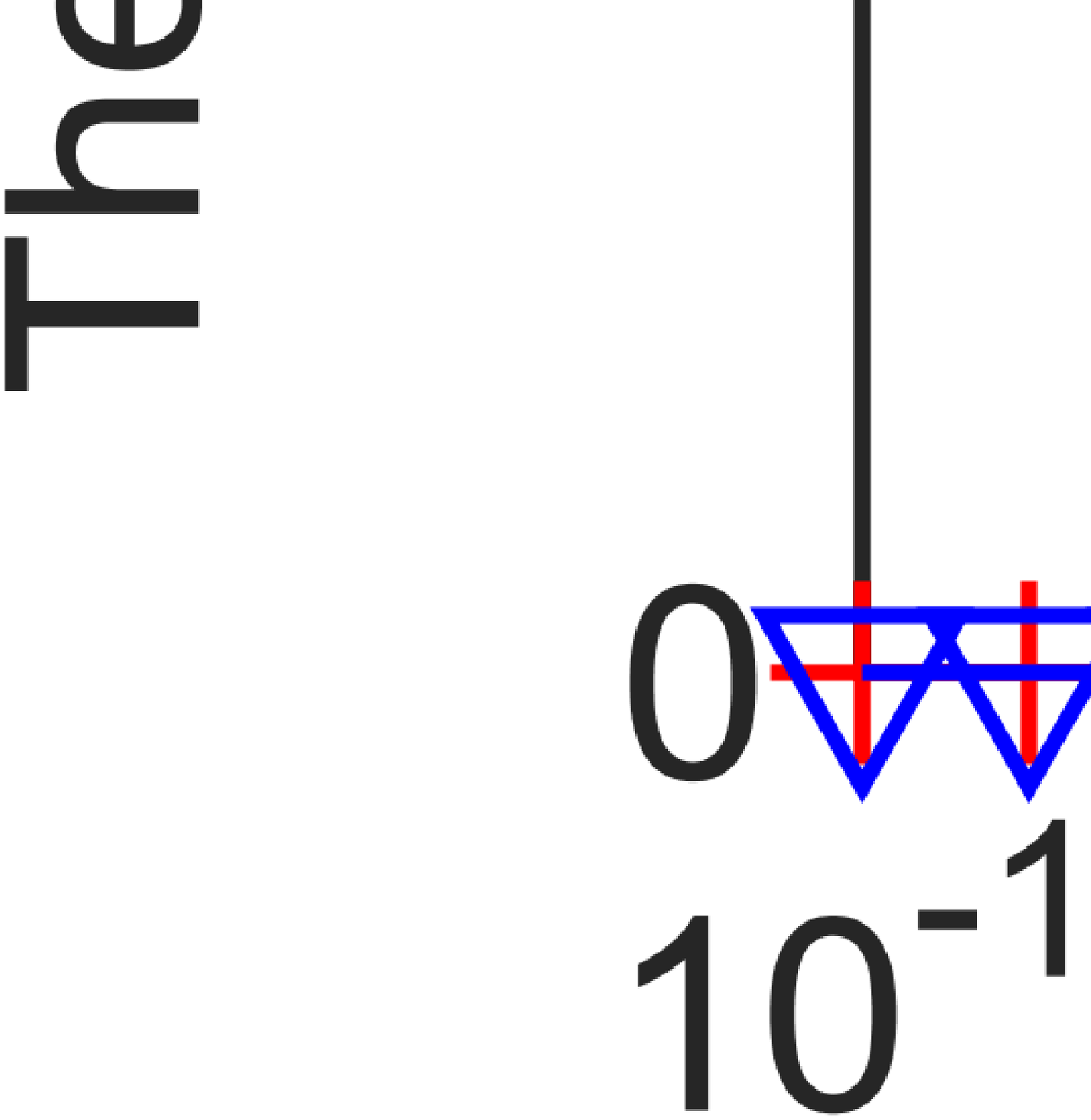}\vspace{-0.1cm}
}\renewcommand{\figurename}{Fig.}\caption{\label{fig:perfm_simp_0dB}Performance results of the 3GPP Case (Rayleigh
fading, no shadow fading) with $\gamma=0\,\mathrm{dB}$.}
\par\end{centering}
\vspace{-0.3cm}
\end{figure*}

\begin{figure*}
\noindent \begin{centering}
\center\subfloat[\label{fig:lambda_tilde_simp_10dB}The active BS density $\tilde{\lambda}$.]{\includegraphics[width=5cm]{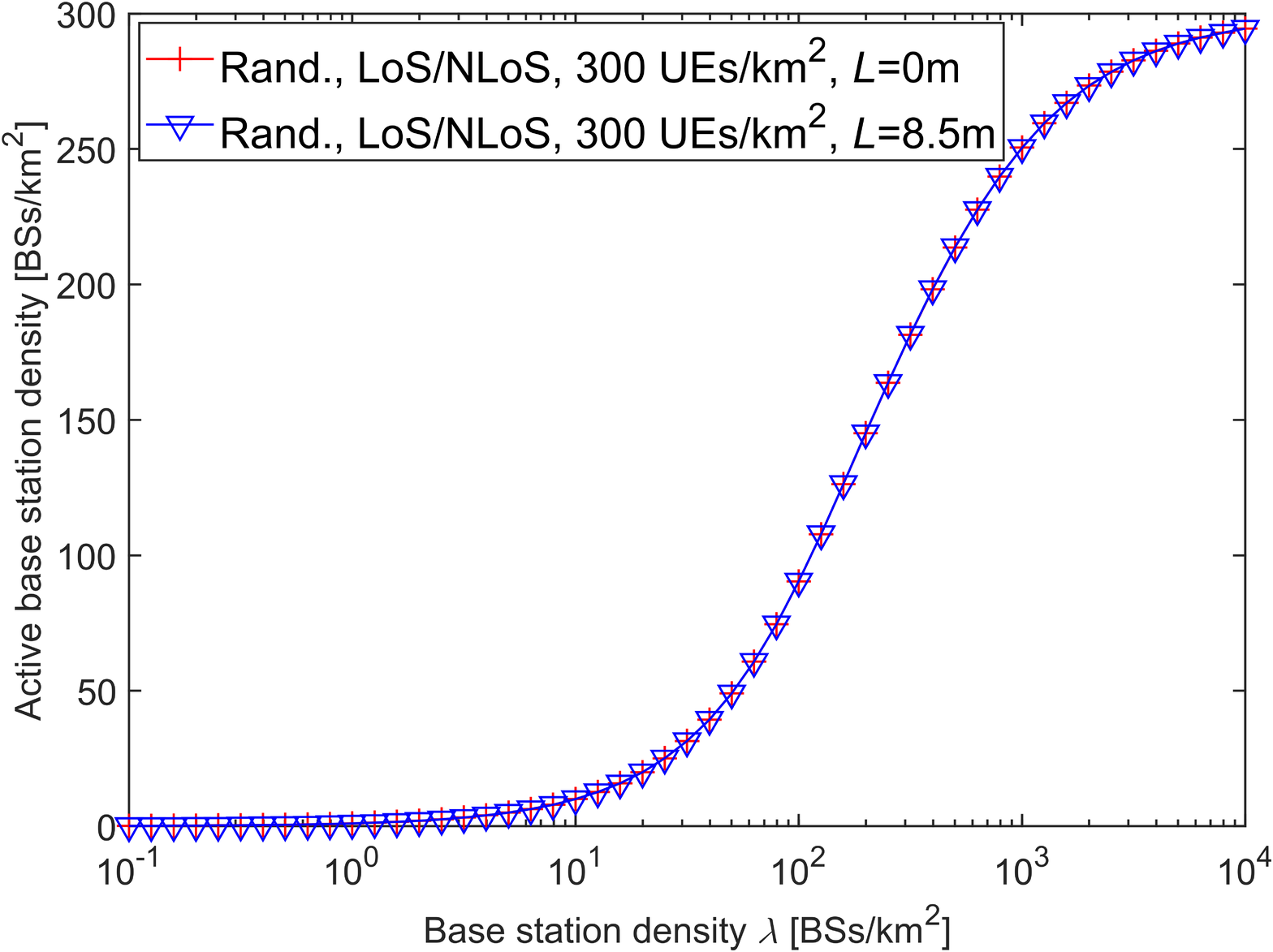}\vspace{-0.1cm}
}$\quad$\subfloat[\label{fig:Pr_cov_simp_10dB}The coverage probability $p^{{\rm {cov}}}\left(\lambda,\gamma\right)$.]{\includegraphics[width=5cm]{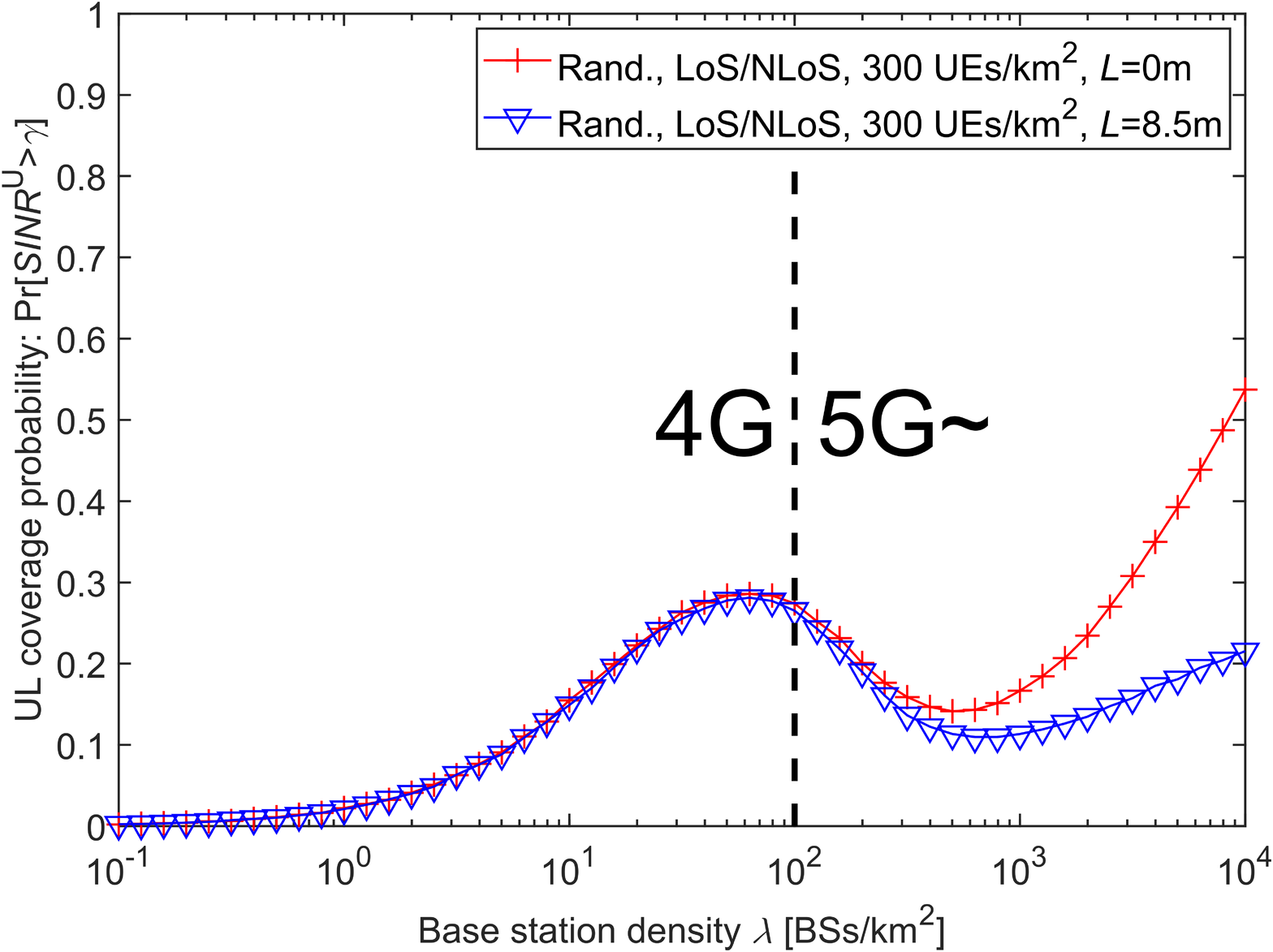}\vspace{-0.1cm}
}$\quad$\subfloat[\label{fig:reliable_UEden_simp_10dB}The density of reliably working
UEs $\tilde{\rho}$.]{\includegraphics[width=5cm]{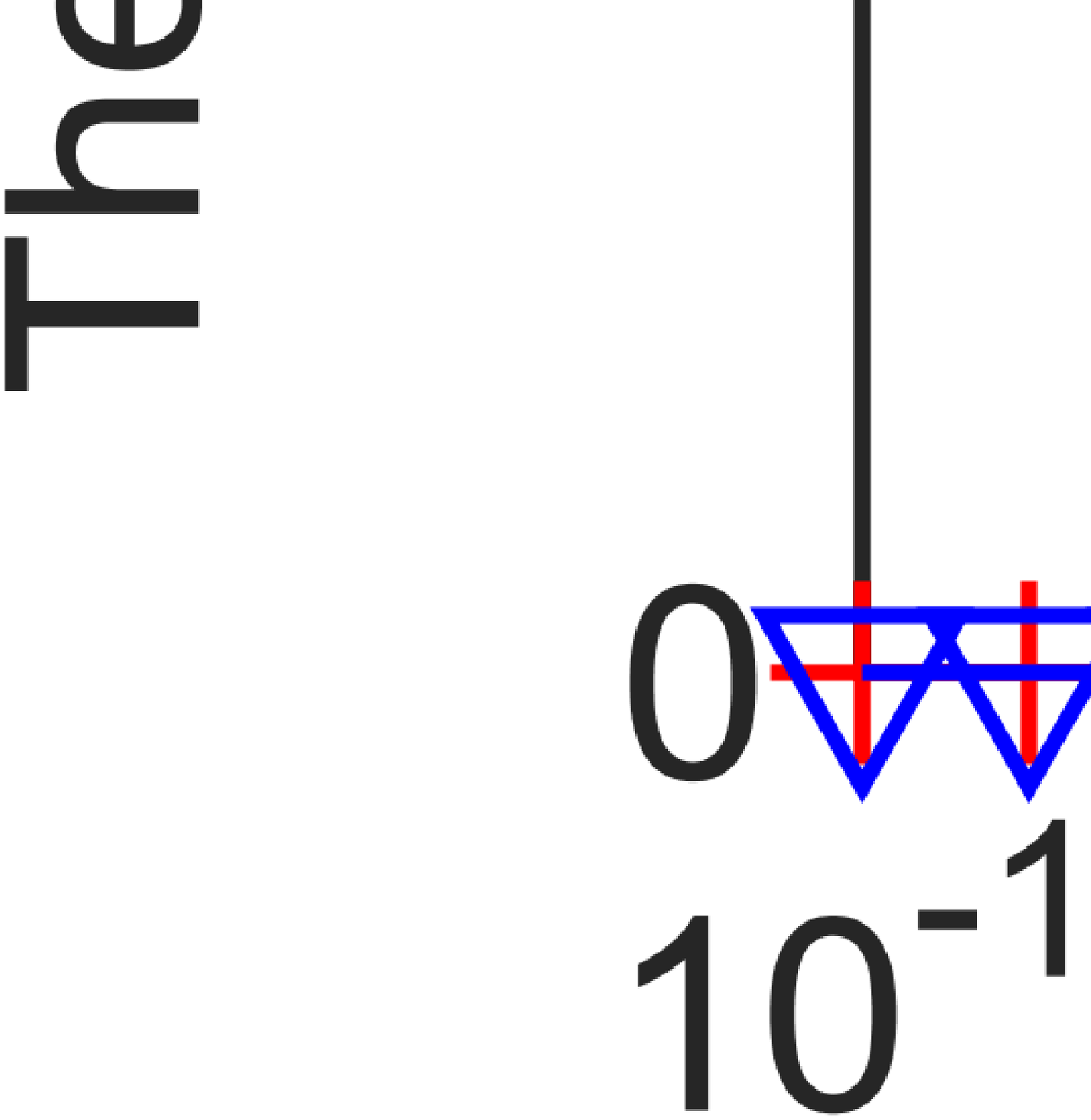}\vspace{-0.1cm}
}\renewcommand{\figurename}{Fig.}\caption{\label{fig:perfm_simp_10dB}Performance results of the 3GPP Case (Rayleigh
fading, no shadow fading) with $\gamma=10\,\mathrm{dB}$.}
\par\end{centering}
\vspace{-0.3cm}
\end{figure*}

In Figs.~\ref{fig:perfm_simp_0dB} and \ref{fig:perfm_simp_10dB},
we plot the performance results of the 3GPP Case for $\gamma=0\,\mathrm{dB}$
and $\gamma=10\,\mathrm{dB}$, respectively.%
{} Here, we only consider realistic networks with the random deployment
of BSs. 
\begin{figure*}
\noindent \begin{centering}
\center\subfloat[\label{fig:lambda_tilde_adv_0dB}The active BS density $\tilde{\lambda}$.]{\includegraphics[width=5cm]{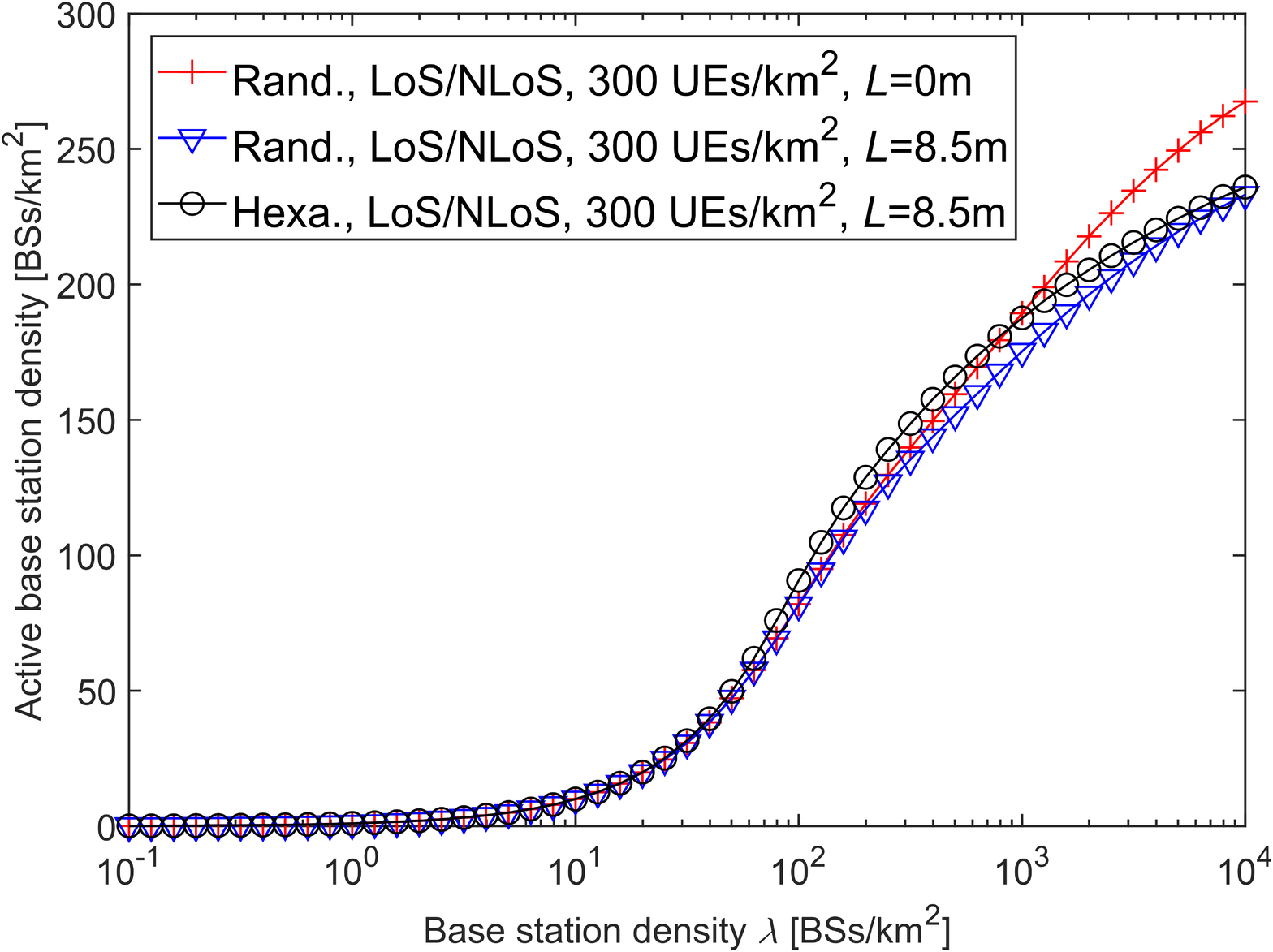}\vspace{-0.1cm}
}$\quad$\subfloat[\label{fig:Pr_cov_adv_0dB}The coverage probability $p^{{\rm {cov}}}\left(\lambda,\gamma\right)$.]{\includegraphics[width=5cm]{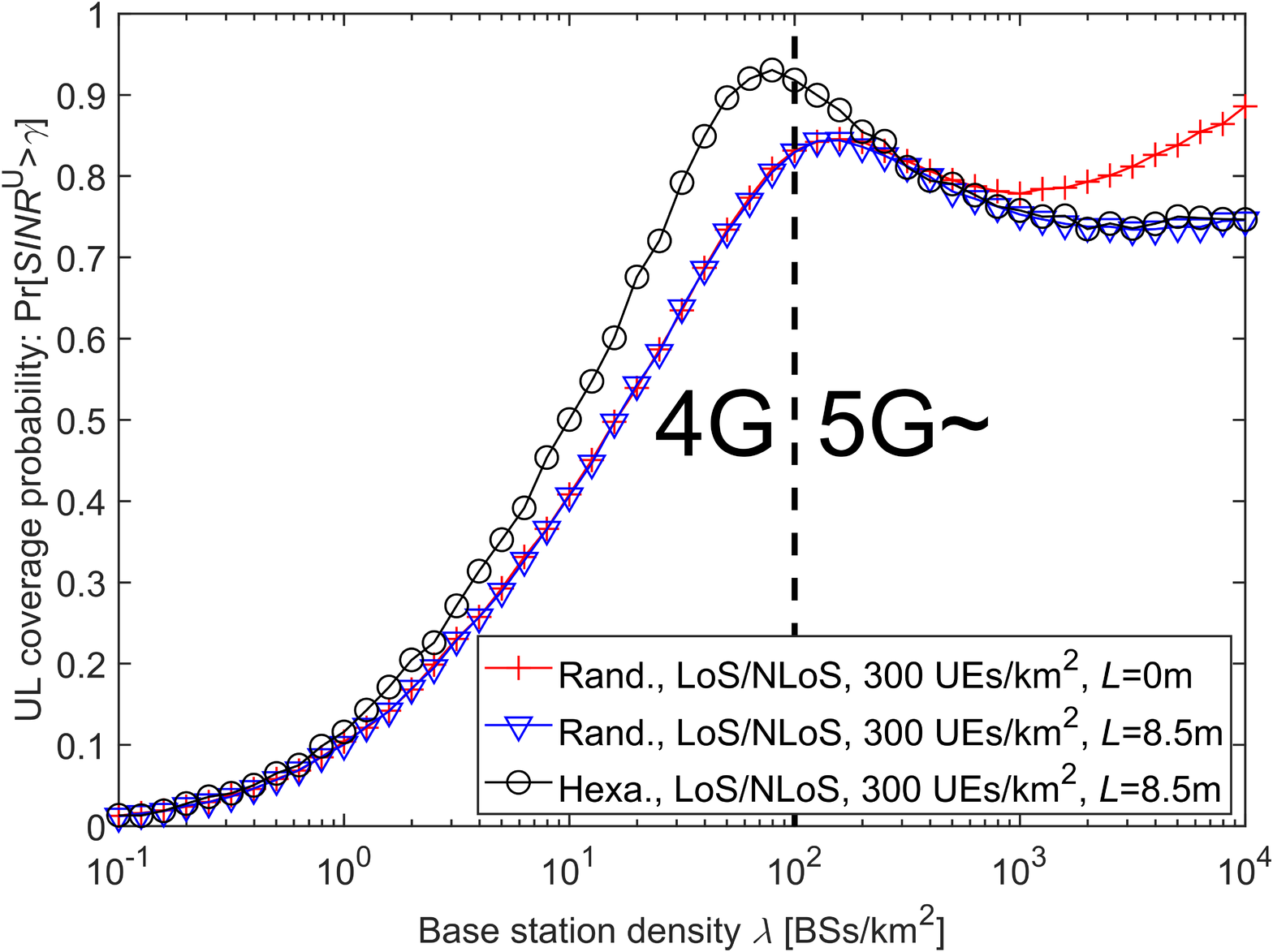}\vspace{-0.1cm}
}$\quad$\subfloat[\label{fig:reliable_UEden_adv_0dB}The density of reliably working
UEs $\tilde{\rho}$.]{\includegraphics[width=5cm]{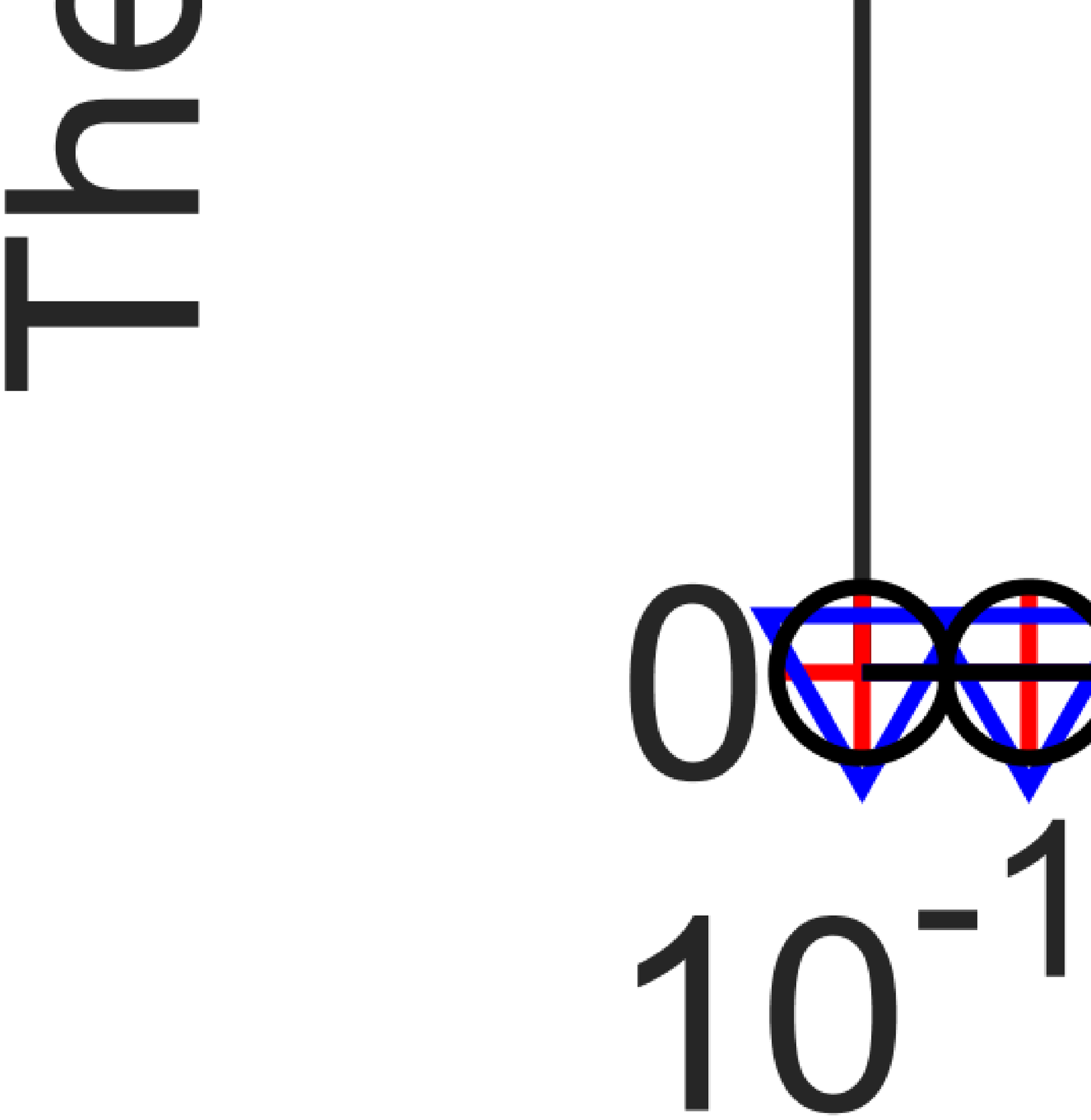}\vspace{-0.1cm}
}\renewcommand{\figurename}{Fig.}\caption{\label{fig:perfm_adv_0dB}Performance results of the Advanced 3GPP
Case (Rician fading for LoS, correlated shadow fading) with $\gamma=0\,\mathrm{dB}$.}
\par\end{centering}
\vspace{-0.3cm}
\end{figure*}
\begin{figure*}
\noindent \begin{centering}
\center\subfloat[\label{fig:lambda_tilde_adv_10dB}The active BS density $\tilde{\lambda}$.]{\includegraphics[width=5cm]{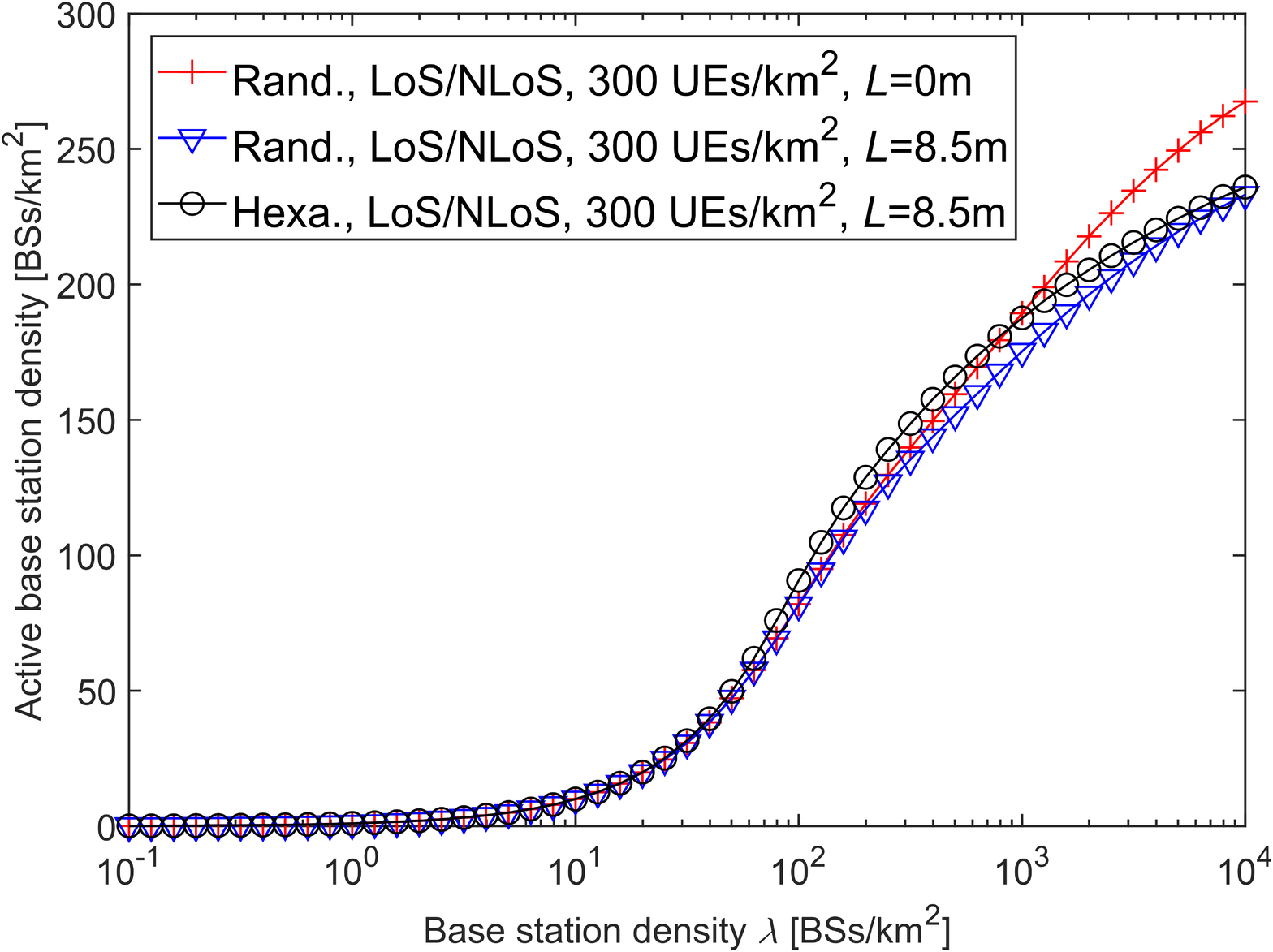}\vspace{-0.1cm}
}$\quad$\subfloat[\label{fig:Pr_cov_adv_10dB}The coverage probability $p^{{\rm {cov}}}\left(\lambda,\gamma\right)$.]{\includegraphics[width=5cm]{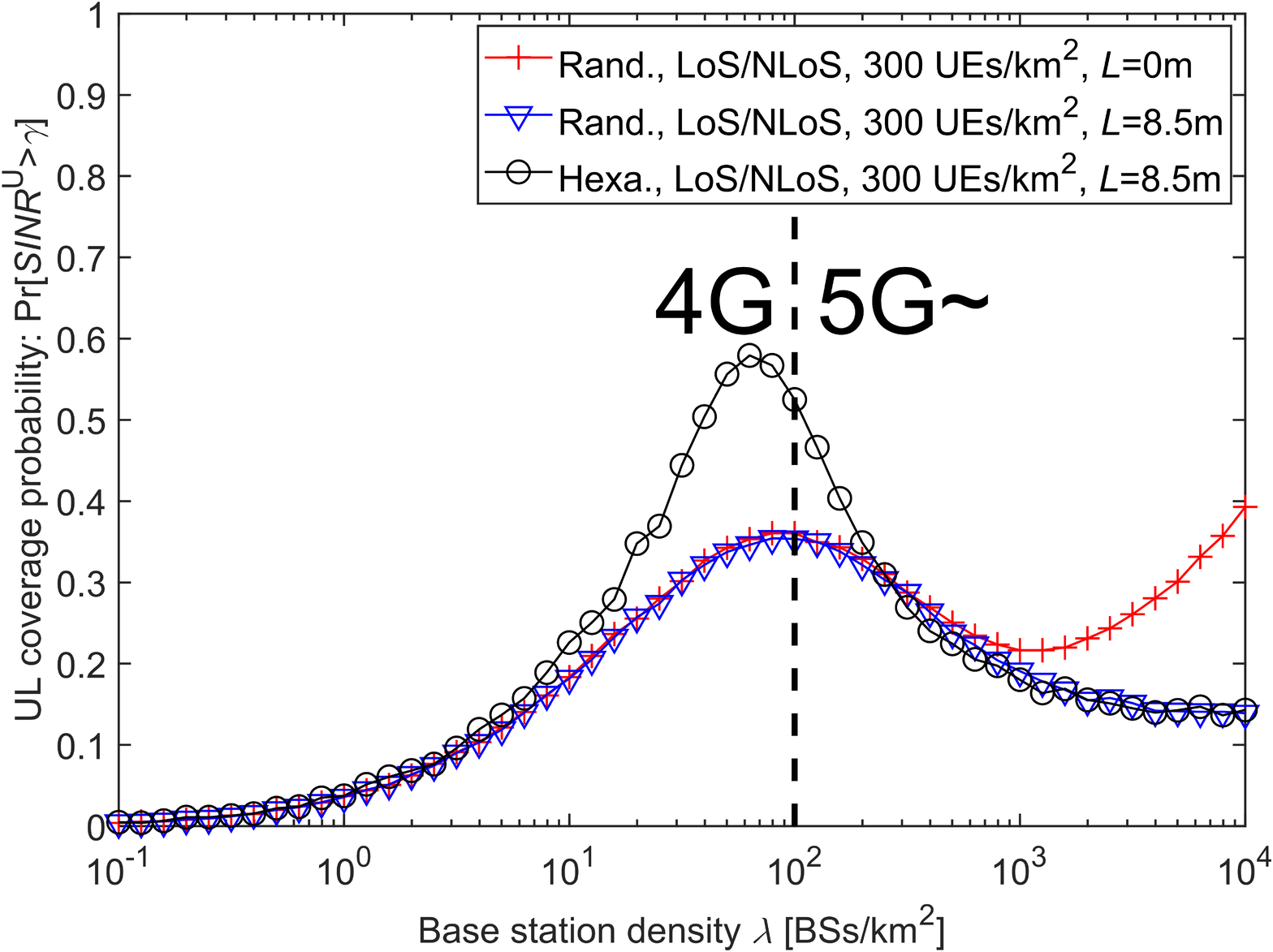}\vspace{-0.1cm}
}$\quad$\subfloat[\label{fig:reliable_UEden_adv_10dB}The density of reliably working
UEs $\tilde{\rho}$.]{\includegraphics[width=5cm]{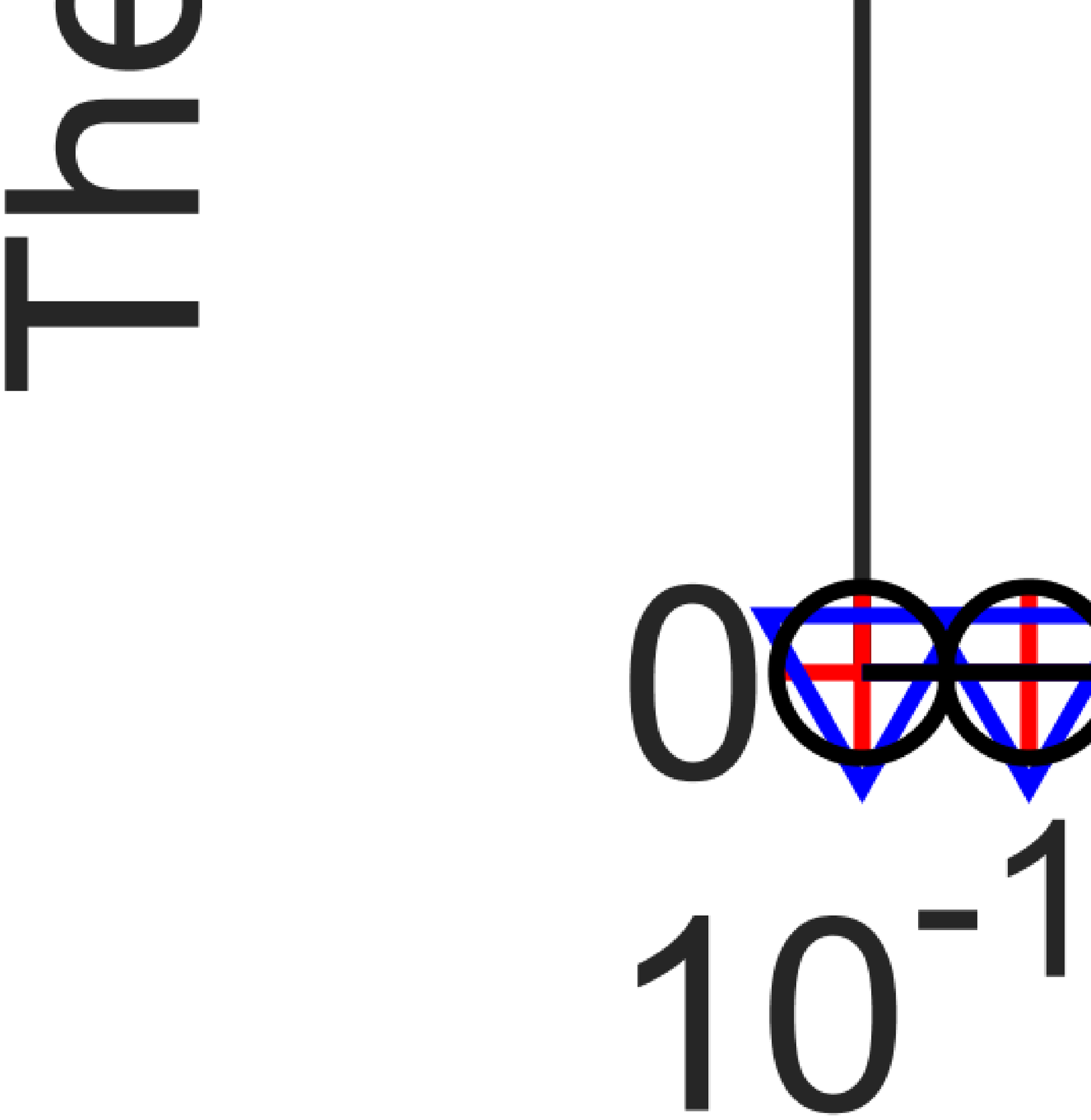}\vspace{-0.1cm}
}\renewcommand{\figurename}{Fig.}\caption{\label{fig:perfm_adv_10dB}Performance results of the Advanced 3GPP
Case (Rician fading for LoS, correlated shadow fading) with $\gamma=10\,\mathrm{dB}$.}
\par\end{centering}
\vspace{-0.3cm}
\end{figure*}
 From these two figures, we can draw the following observations:
\begin{itemize}
\item Figs.~\ref{fig:lambda_tilde_simp_0dB} and~\ref{fig:lambda_tilde_simp_10dB}
show that the active BS density $\tilde{\lambda}$ monotonically increases
with the network densification, and it is bounded by $\rho=300\thinspace\textrm{UEs/km}^{2}$.
Such results are in line with the analytical results in (\ref{eq:lambda_tilde_Huang})~\cite{dynOnOff_Huang2012,Ding2016TWC_IMC}.
However, the density of reliably working UEs $\tilde{\rho}$, i.e.,
$\tilde{\lambda}p^{{\rm {cov}}}\left(\lambda,\gamma\right)$ defined
in (\ref{eq:reliable_UE_den_def}), does not necessarily grow as the
BS density $\lambda$ increases. This is because $p^{{\rm {cov}}}\left(\lambda,\gamma\right)$
is a non-monotone function with respect to $\lambda$, which will
be explained in the following. 
\item When the BS density $\lambda$ is around $\lambda\in\left[10^{-1},70\right]\,\textrm{BSs/km}^{2}$,
the network is noise-limited, and thus $p^{\textrm{cov}}\left(\lambda,\gamma\right)$
increases with $\lambda$ as the network is lightened up with coverage
and the signal power strength benefits form LoS transmissions. 
\item When the BS density $\lambda$ is around $\lambda\in\left[70,400\right]\,\textrm{BSs/km}^{2}$,
$p^{\textrm{cov}}\left(\lambda,\gamma\right)$ decreases with $\lambda$.
This is due to the transition of a large number of interfering paths
from NLoS to LoS, which accelerates the growth of the aggregate inter-cell
interference. Such performance behavior has been reported in \cite{our_work_TWC2016}
for the DL and~\cite{Ding2017ULlos} for the UL.
\item When $\lambda\in\left[400,10^{4}\right]\,\textrm{BSs/km}^{2}$, $p^{\textrm{cov}}\left(\lambda,\gamma\right)$
continuously increases thanks to the BS IMC~\cite{Ding2016TWC_IMC},
i.e., the signal power continues increasing with the network densification
due to the BS diversity gain, while the aggregate interference power
becomes constant, as $\tilde{\lambda}$ is bounded by $\rho$. However,
as shown in Figs.~\ref{fig:Pr_cov_simp_0dB} and~\ref{fig:Pr_cov_simp_10dB},
the antenna height difference $L$ between BSs and UEs has a large
impact on $p^{\textrm{cov}}\left(\lambda,\gamma\right)$ because a
non-zero $L$ places a bound on the signal power strength, which degrades
the coverage performance. In more detail, when $\lambda=10^{4}\,\textrm{BSs/km}^{2}$
and $\gamma=0\,\mathrm{dB}$, the coverage probability with $L=8.5\,\mathrm{m}$
loses 13$\,$\% compared to that with $L=0\,\mathrm{m}$. Such performance
degradation further enlarges to 32$\,$\% when $\lambda=10^{4}\,\textrm{BSs/km}^{2}$
and $\gamma=10\,\mathrm{dB}$, showing a much less chance of UE working
reliably above $10\,\mathrm{dB}$. 
\item Due to the complicated performance behavior of $p^{\textrm{cov}}\left(\lambda,\gamma\right)$,
the density of reliably working UEs $\tilde{\rho}$ displayed in Figs.~\ref{fig:reliable_UEden_simp_0dB}
and~\ref{fig:reliable_UEden_simp_10dB} depends on the following
factors:
\begin{itemize}
\item For a low-reliability criterion, such as surpassing a UL SINR threshold
of $\gamma=0\,\mathrm{dB}$, the density of reliably working UEs $\tilde{\rho}$
grows quickly with the network densification, showing the benefits
of UL IoT UDNs. In contrast, for a high-reliability criterion, such
as surpassing a UL SINR threshold of $\gamma=10\,\mathrm{dB}$, the
density of reliably working UEs $\tilde{\rho}$ does not exhibit a
satisfactory performance even in UDNs, e.g., we merely get $\tilde{\rho}<50\,\textrm{UEs/km}^{2}$
when $\lambda=10^{3}\,\textrm{BSs/km}^{2}$. The situation only improves
when the BS IMC fully kicks in, e.g., $\lambda>10^{3}\,\textrm{BSs/km}^{2}$. 
\item Considering the existence of a non-zero $L$, the density of reliably
working UEs $\tilde{\rho}$ could even \emph{decrease} as we deploy
more BSs (see Fig.~\ref{fig:reliable_UEden_simp_10dB}, $\lambda\in\left[200,600\right]\,\textrm{BSs/km}^{2}$).
This calls for the usage of sophisticated interference management
schemes~\cite{Book_CoMP} in UL IoT UDNs. Another solution to mitigate
such strong inter-cell interference is beam steering/shaping using
multi-antenna technologies~\cite{Tutor_smallcell}. 
\end{itemize}
\end{itemize}

\subsection{Performance Results of the Advanced 3GPP Case\label{subsec:Performance-Results-3GPP-advCase}}

In Figs.~\ref{fig:perfm_adv_0dB} and \ref{fig:perfm_adv_10dB},
we plot the performance results of the Advanced 3GPP Case for $\gamma=0\,\mathrm{dB}$
and $\gamma=10\,\mathrm{dB}$, respectively. To make our study more
complete, here we consider both the random and the hexagonal deployments
of BSs. From these two figures, we can see that the previous conclusions
are qualitatively correct, which indicates that it is not urgent to
investigate Rician fading and/or correlated shadow fading in the context
of UDNs. However, there are two new observations that are worth mentioning: 
\begin{itemize}
\item From Figs.~\ref{fig:lambda_tilde_simp_0dB}/\ref{fig:lambda_tilde_simp_10dB}
and Figs~\ref{fig:lambda_tilde_adv_0dB}/\ref{fig:lambda_tilde_adv_10dB},
we can see that the active BS density $\tilde{\lambda}$ of the Advanced
3GPP Case is smaller than that of the 3GPP Case. This means that (\ref{eq:lambda_tilde_Huang})
is no longer accurate to characterize $\tilde{\lambda}$ for the Advanced
3GPP Case. This is because the correlated shadow fading allows a BS
with a lower environmental fading factor, i.e., $S_{b}^{{\rm {BS}}}$,
to attract more UEs than other BSs with higher values of $S_{b}^{{\rm {BS}}}$.
Its theoretical analysis is an open problem for further study. 
\item The hexagonal deployment of BSs can improve network performance for
relatively sparse networks (e.g., $\lambda<10^{2}\,\textrm{BSs/km}^{2}$),
but not for UDNs (e.g., $\lambda>10^{3}\,\textrm{BSs/km}^{2}$). Such
conclusion indicates that it is not urgent to investigate the performance
of UDNs with the hexagonal deployment of BSs, which has been a long-standing
open problem for decades~\cite{Our_DNA_work_TWC15}. 
\end{itemize}

\section{Conclusion\label{sec:Conclusion}}

We presented simulation results to evaluate the network performance
of UL IoT UDNs. From our study, we can see that for a low-reliability
criterion, the density of reliably working UEs grows quickly with
the network densification. However, in our journey to realize a more
reliable UL IoT UDNs, we should be aware of several caveats:
\begin{itemize}
\item First, for a high-reliability criterion, the density of reliably working
UEs remains low in UDNs due to excessive inter-cell interference,
which should be considered when operating UL IoT UDNs. 
\item Second, due to the existence of a non-zero antenna height difference
between BSs and UEs, the density of reliably working UEs could even
\emph{decrease} as we deploy more BSs. This calls for further study
of UL IoT UDNs. 
\item Third, well-planned hexagonal-like BS deployments can improve network
performance for relatively sparse networks, but not for UDNs, showing
that alternative solutions other than BS position optimization should
be considered in the future UL IoT UDNs.
\end{itemize}
\bibliographystyle{IEEEtran}
\bibliography{Ming_library}

\end{document}